\documentclass[smallextended]{svjour3}

\usepackage{amsmath}
\usepackage{amssymb}
\usepackage{graphicx}
\usepackage{subfig}

\begin{document}
\newcommand{\qd}{\hbox{$^{\prime\prime}$\kern-.15cm{,}\kern.04cm}}
\newcommand{\arc}{\hbox{$^{\prime\prime}$}}
\newcommand{\eq}[1]{(\ref{#1})}
\newcommand{\rcirc}{\hbox{$^{\circ}$}}

\title{\textit{Software correlator for Radioastron mission}}

\author{Sergey F.~Likhachev \and Vladimir I.~Kostenko \and Igor A.~Girin \and Andrey S.~Andrianov \and Vladimir E.~Zharov \and Alexey G. Rudnitskiy}

\institute{Sergey F.~Likhachev \and Vladimir I.~Kostenko \and Igor A.~Girin \and Andrey S.~Andrianov \and Alexey G. Rudnitskiy
\at
Astro Space Center of Lebedev Physical Institute of RAS,\\ 84/32 Profsoyuznaya st., Moscow, 117997, Russia\\\email{vkostenko@asc.rssi.ru}
\and
Vladimir E.~Zharov
\at
Lomonosov Moscow State University, Physics Faculty, Sternberg Astronomical Institute,\\
Universitetskij prospect, 13, Moscow, 119991, Russia\\\email{VladZh2007@yandex.ru}}

\maketitle

\begin{abstract}
\textit{In this paper we discuss the characteristics and operation of Astro Space Center (ASC) software FX correlator that is an important component of space-ground interferometer for Radioastron project. This project performs joint observations of compact radio sources using 10 meter space radio telescope (SRT) together with ground radio telescopes at 92, 18, 6 and 1.3 cm wavelengths. In this paper we describe the main features of space-ground VLBI data processing of Radioastron project using ASC correlator. Quality of implemented fringe search procedure provides positive results without significant losses in correlated amplitude. ASC Correlator has a computational power close to real time operation. The correlator has a number of processing modes: ``Continuum'', ``Spectral Line'', ``Pulsars'', ``Giant Pulses'',``Coherent''. Special attention is paid to peculiarities of Radioastron space-ground VLBI data processing. The algorithms of time delay and delay rate calculation are also discussed, which is a matter of principle for data correlation of space-ground interferometers. During 5 years of Radioastron space radio telescope (SRT) successful operation, ASC correlator showed high potential of satisfying steady growing needs of current and future ground and space VLBI science. Results of ASC software correlator operation are demonstrated.}

\end{abstract}



\section{Introduction}~\label{intro}

Very Long Baseline Interferometry (VLBI) allows to reach the highest angular resolution available in radio astronomy for detailed study of the most distant objects in the Universe. Targets of interest include active galactic nuclei (AGN), radio galaxies, masers and pulsars. Studying radio emission from these objects with high angular resolution provide unique insight into the physics of these objects. For general overview of VLBI development and historical evolution of radio interferometry, see~\cite{Thompson}.

Introduction of space element in VLBI extends the potential of this method. Space radio telescope (SRT) operating in elliptical orbit around the Earth observe the same source as ground radio telescopes. As a result the synthesized aperture has a size much larger than the Earth diameter enabling to obtain images of the most compact sources with very high angular resolution (up to a few microarcseconds).

The first VLBI experiments using space telescope were conducted from 1986 to 1988 with a 4.9-meter antenna onboard of a geostationary satellite~\cite{Levy86,Linfield}, which was a part of NASA Tracking and Data Relay Satellite System. Successful observations were carried out at frequencies \textbf{of} 2.3 and 15 GHz, demonstrating the possibilities of space-ground VLBI. It was shown that non-standard VLBI procedures were necessary to obtain \textbf{an} adequate phase stability and to correlate the data~\cite{Levy89}.

This experience was used in HALCA (Highly Advanced Laboratory for Communications and Astronomy) space satellite of VLBI Space Observatory Program (VSOP)~\cite{Hira,Hirosawa}. HALCA was developed by Japanese Institute of Space and Astronautical Science (ISAS). The satellite was launched in February 1997 and had an 8-meter onboard radio telescope that operated together with ground telescopes to obtain images of radio sources at 1.6 and 5 GHz. The following correlators were used to process the data of VSOP mission: VSOP Correlator at Mitaka, VLBA at Socorro and S2 at Penticton.

Space-ground interferometer implemented in Radioastron mission~\cite{Kardashev2009,Kardashev2013} is aimed to study physical parameters and angular structure of galactic and extra galactic radio sources using VLBI techniques. The main component of Radioastron mission is the space satellite ``Spektr-R'' operating in elliptical orbit around the Earth. Short description of the space radio telescope is given in Section~\ref{SRT}.

In 2007 - 2009 a software FX correlator was developed in Astro Space Center of Lebedev Physical Institute of Russian Academy of Sciences (ASC LPI) in order to support data processing of Radioastron mission. It was tested using the data obtained from ground interferometers. Results of ASC correlator tests showed that they are similar to DiFX correlator -- a software correlator that is used by many VLBI arrays including VLBA, EVN, etc.~\cite{Deller}. New correlation modes for space-ground interferometer data were added in 2010 and first implemented in November 2011, four months later after the launch of ``Spektr-R'' space satellite.

Fundamentals and specific features of VLBI data processing for space-ground interferometry are described in Section~\ref{VLBI1}.

The main task of a correlator is to perform a coherent digital combination of independently received signals from a pair of interferometer antennas. Correlator output consists of complex visibilities calculated for each baseline. Complex visibilities represent spatial spectral components of the source brightness distribution.

For Radioastron mission a hardware correlator (e. g. VSOP hardware correlator) wasn't considered, because the software correlator is more flexible and adjustable during the operation. It has a simple interface and architecture that allows to perform the correlation of different data formats, use the new models of geometric delay and change correlation mode. During the first two years of the mission operation, regular updates of ASC correlator were performed in order to meet all upcoming observational tasks.

Space-ground baselines that are implemented in Radioastron mission allow to propose a number of typical strategies of observations and scientific goals:
\begin{itemize}
    \item survey observations to measure the angular structure and brightness temperatures of compact extragalactic radio sources,
    \item imaging of AGN, radio galaxies, maser sources, pulsars and studying of
their structure and evolution,
    \item astrometric measurements.
\end{itemize}

Each of strategies support different type of scheduling and different algorithms of correlation procedures. Sources can be observed during
a few hours by data records (``scans'') each 5 -- 10 minutes in duration, which is specified by Primary Investigator in the proposal. Depending on the strategy ASC correlator can operate in different modes.

In this paper we pay special attention to the methods of time delay and delay rate computing for space-ground interferometer as well as to the differences between space-ground and ground VLBI. Description of data processing procedure is given in Section~\ref{Reduction}.

\section{The space radio telescope}~\label{SRT}

``Spektr-R'' is the space radio telescope for Radioastron mission. It is operating in a perturbed orbit. The orbital parameters were chosen to maximize their evolution using weak gravitational perturbations from the Moon and the Sun. Perigee of orbit varies from 400 to 65,000 km, apogee varies from 265,000 to 360,000 km, the eccentricity of the orbit changes from 0.59 to 0.96 and initial inclination of orbital plane is $51\rcirc$. Orbital period of the space radio telescope is about 9 days.

The space radio telescope has a 10-m parabolic antenna. The surface accuracy (RMS) of the antenna is $\pm0.5$ mm. Obervational frequency bands are: 0.327 GHz (P-band), 1.665 GHz (L-band), 4.830 GHz~(C-band)~and 18.392-25.112 GHz (K-band). Receivers at each frequency band have two independent channels for right-hand and left-hand circular polarization (RCP and LCP). Bandwidth is 16 MHz for P-band receiver and 32 MHz for L-, C-, and K-band receivers correspondingly. The detailed information of the space radio telescope specifications and scientific programs can be found in ~\cite{RA1,RA2}. The SRT has an onboard active hydrogen maser frequency standard (H-maser) giving references for frequency conversions and data sampler.

The onboard digital receiver backend samples the analog signal with one or two-bit sampling. The samples are taken at intervals of $1/(2\Delta f)$ where $\Delta f$ is a channel bandwidth. The digital backend generates four binary data streams corresponding to two polarization and two (USB and LSB) IF frequency channels. Digitized VLBI data streams are transmitted from the satellite down to the ground tracking stations as a series of frames. Each frame has a header that contains telemetry information of the space radio telescope status.

The space radio telescope with onboard H-maser has no frequency counter or clock. During the observations data is synchronized by onboard frequency standard and transmitted to the tracking station. At the ground tracking station synchronization sequence and data extracted in demodulator and decoder and directed to Radioastron Data Recorder (RDR) HDD system. Data sequence counter resets before the start of new observing session and wait start of ``write'' command (at scheduled UTC second) from station clock governed by local H-maser. RDR recording starts upon arrival of first data frame right in the scheduled UTC second.  After that synchronization of incoming onboard data flow will be supported continuously till the end of observing session. The time offset between ground station and onboard clock can be established by restoring of time delay between Radioastron and tracking station from the ballistic prediction of the space radio telescope position.

The primary tracking station for Radioastron project is a 22-m radio telescope of Pushchino Radio Astronomical Observatory (PRAO). The secondary tracking station is located in Green Bank (43-m antenna, NRAO, USA). It tracks and receives the data in the Western hemisphere. Tracking stations follow the spacecraft across the sky and receive digital data from the space radio telescope via downlink at 15 GHz carrier frequency with a maximum rate of 144 Mbps followed by recording it locally to the disks using RDR. Uplink signal at 8.4 GHz can be used as a frequency/time standard as described in Section 4.1 (``Coherent'' mode). Data from tracking stations is transferred to ASC correlator via broadband network connection~\cite{Kardashev2013}.

\section{Fundamentals of VLBI technique: delay compensation and fringe stopping}\label{VLBI1}

Below we illustrate the concepts of two-element interferometer. Correlator output in complex form can be represented as:
\begin{equation}
	\label{a1}
	r(\tau) = \langle V_1(t)\cdot V_2^*(t-\tau)\rangle = \lim\limits_{T\rightarrow\infty}\frac{1}{T}\int\limits_{-T/2}^{T/2}
	V_1(t)\cdot V_2^*(t-\tau)dt,
\end{equation}
where $V_1, V_2$ correspond to the signal voltage measured from telescopes ``1'' and ``2'', ``*'' represent complex conjugation.

Delay $\tau$ in Eq. (\ref{a1}) include geometrical delay $\tau_g$ and corrections that depend on the effects of radio wave propagation, as well as on the uncertainties in the clock offset on VLBI sites.

The delay $\tau_g$ for ground VLBI changes slowly in time due to the rotation of the Earth. For space-ground interferometry the changes in the delay $\tau_g$ in time are caused not only by the rotation of the Earth, but also by the space radio telescope orbital motion.

%Delay is tracked in the correlator step by step with the sample period $T_{sample}$ equal to Nyquist rate $\eta=1/(2\Delta f)$ where $\Delta f$ is receiver bandwidth. The difference between ``a priori'' delay and the quantized delay during the correlation procedure can be $\pm 0.5$ of sample period, because the delay can be tracked only for the nearest single sample. This difference vary in time due to the presence of delay rate. 

The measured residuals between ``a priori'' delay and ``real'' can be expressed as:

\begin{equation}
\label{a3}
	\tau(t) = \tau(t_0) +
	[\dot{\tau}(t_0)+\Delta\dot{\tau}(t_0)](t-t_0) +
	\frac{1}{2}[\ddot{\tau}(t_0)+\Delta\ddot{\tau}(t_0)](t-t_0)^2
	+\cdots,
\end{equation}
where $t_0$ is the beginning of scan.

Terms $\tau(t_0), \dot{\tau}(t_0), \ldots$ can be calculated using ``a priori'' parameters of interferometer. Terms $\Delta\dot{\tau} (t-t_0)$, $ \frac{1}{2}\Delta\ddot{\tau} (t-t_0)^2$ arising from uncertainty in delay model $\tau$ should not be greater than one sample period $T_{sample}=1/(2\Delta f)$ during the total integration time $T_{INT}$:

\begin{equation}
	\label{a4}
	\Delta\dot{\tau}\cdot T_{INT} +
	\frac{1}{2}\Delta\ddot{\tau}\cdot T_{INT}^2<T_{sample}.
\end{equation}

For space-ground baselines the terms $\Delta\dot{\tau}, \Delta\ddot{\tau}$ contain contribution both from onboard H-maser frequency standard instability and errors in the reconstructed velocity and acceleration of the SRT. It is the main difference between the correlation procedures for space-ground and ground baselines, where the motion of ground telescopes is known precisely.

Below we estimate the contribution of various factors related to the SRT orbit affecting on delay and delay rate. These factors are important, because they determine the requirements for correlator.

The maximum value of geometrical delay $\tau_g$ and delay rate $\dot{\tau_g}$ for ground VLBI is about $21$ $\mu$s and $\leq 3$ $\mu$s $\cdot$s$^{-1}$ correspondingly. Comparison of this value with the sampling period, which is equal to $0.03$ $\mu$s ($30$ ns) for bandwidth $\Delta f = 16$ MHz, showed that delay must be updated as frequent as 100 times per second.

Positions of ground telescopes are known with high accuracy (few centimeters) leading to the errors of $\tau(t_0)$ in \eq{a3} of $\sim30$ ps. It is significantly less than the value of sampling period ($30$ ns) for $\Delta f = 16$ MHz bandwidth. The situation is different for the space telescope. The position of the SRT is calculated by Ballistics Center of Keldysh Institute of Applied Mathematices of Russian Academy of Sciences (KIAM). They use the theoretical model of forces acting on the space radio telescope. This model provides an accuracy of 200 m in the position corresponding to the delay errors of 0.7 $\mu$s. Real measurements of the SRT position can exceed theoretical accuracy. Clock delays for the SRT can reach values up to 10 $\mu s$. Additionally, delays for certain ground stations aren't known with an adequate accuracy. Thus initial fringe search should be performed in a wider correlation window with at least 2048 sampling periods ($\pm$ 64 $\mu$s).

Traditional delay model used for ground baselines in case of space-ground interferometer doesn't consider factors related to the SRT orbit. This is one of the main reasons to revise the traditional VLBI delay model used for ground telescopes~\cite{IERS2010}. Delay rate for the space radio telescope operating in elliptical orbit around the Earth is $\dot{\tau}(t_0) = \dot{\overrightarrow{b}} \cdot \overrightarrow{s}/c$. It depends not only on the relative velocity of the telescopes $\dot{\overrightarrow{b}}$ but also on the drift of the onboard H-maser frequency standard. This drift can be explained by motion of the SRT, relativistic effects (see \cite{RA1}) and properties of the onboard H-maser. It can be order of $30-35$ $\mu$s/s$^{-1}$.

Maximum delay acceleration for ground VLBI is about $110$ ps/s$^2$. Effect of acceleration for 2 s integration time will be about 220 $ps/s^2$ and thus can be neglected. Acceleration of the SRT near apogee is about $5-6$ m/s$^2$ leading to the delay acceleration of $\sim 20\ \textrm{ns}/ \textrm{s}^2$. According to~\eq{a3} the correction of delay can reach $\sim 40$ ns for 2 s integration time. \textit{For the space radio telescope it is necessary to take into account acceleration in the delay model.}

Differences in delay and delay rate for ground and space-ground configurations are shown in Table ~\ref{tab:0}.

\begin{table}[th]
\caption{Estimated delay and delay rate for ground and space-ground VLBI.}
\label{tab:0}       
\center
\begin{tabular}{|l|c|c|c|}
\hline
				& $\tau_g$, s 			& $\dot{\tau}, \mu$s/s 	& $\ddot{\tau}, ns/s^2$  	\\
\hline
Ground VLBI      		& $21\times 10^{-3}$ 	& 3 					&0.1  				\\
Ground-space VLBI 	& 1 					& $>30$ 				&$>20$  				\\
\hline
\end{tabular}
\end{table}

Fringe search and fringe fitting procedures for space-ground VLBI are different comparing to the similar for ground VLBI. The sensitivity of the SRT is lower than for the ground telescopes. Assuming a 300-second integration, in order to detect fringe, for example the minimal correlated flux density must be $S_{min}\geq$ 112, 21, 35, 91 mJy for P-, L-, C-, K- bands correspondingly for SRT-Green Bank Telescope baseline.

Fringe search procedure for space-ground VLBI can be challenging than for ground VLBI, because the fringe delay and delay rate (position in correlation window) strongly depends on the errors in position and velocity of the SRT, as well as the onboard frequency standard drift. According to \eq{a4} the total integration time $T_{INT}$ is limited by the errors in the SRT velocity. 
In order to get the highest possible sensitivity for space-ground VLBI it is necessary to use the total integration time $T_{INT}=T_c$ (coherence time). Phase fluctuations limiting the coherent time are caused by atmosphere instability and phase noise of frequency standards and receiver frequency conversion elements at the SRT and ground stations~\cite{Thompson,Moran76}. Initially the value of $T_c$ is unknown and one of the goals is to estimate it from the observation. For example, in case the onboard H-maser instability is $\approx 10^{-14}$ s/s the coherent time is equal to 700 s for K-band observations.

\section{Correlator overview.}~\label{Corr}
\subsection{Data correlation. Processing algorithm.}~\label{Proc}

Correlation procedure is the crucial part of Radioastron VLBI data processing. About 95\% of Radioastron observational data is processed with ASC Correlator.

\begin{figure}[th]
\includegraphics[width=\columnwidth]{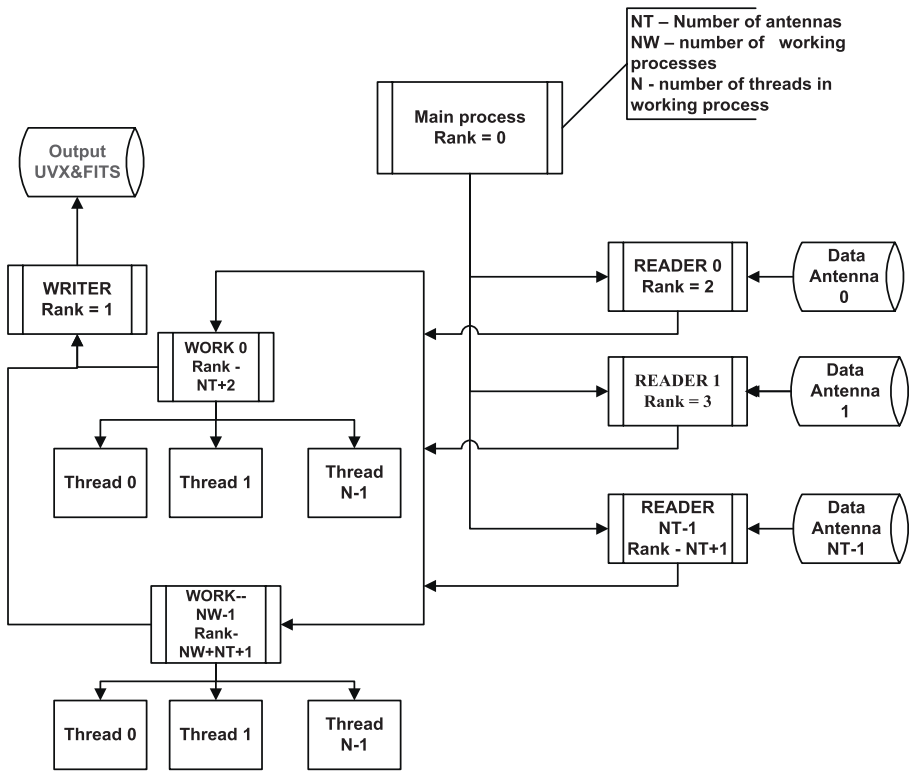}
\caption{\small Structure of ASC correlator multi-threaded architecture.} \label{fig1}
\end{figure}

The main feature of ASC correlator is delay model -- ORBITA2012. Comparison between ASC correlator delay model and DiFX correlator CALC~\cite{Gordon} delay model showed that the difference in delay and fringe rate for ground baselines does not exceed $10^{-13}$ s and $10^{-14}$ s/s correspondingly. ORBITA2012 model perform the calculation of signal delay between the SRT and ground tracking station more accuractely than CALC, since this delay is calculated up to the 2$^{nd}$ order of Taylor series, while CALC model calculate the delay up to the 1$^{st}$ order. This peculiarity of ORBITA2012 delay model allows to calculate the delay for space-ground baselines accurately leading to the decrease of correlation window size for the first correlator pass.

As a software FX-correlator (Fourier transform -- Spectrum multiplication) ASC correlator provides evaluation of cross-correlation functions for multiple VLBI stations for ground and space-ground baselines \cite{Benson}.

In ASC correlator the signal from each telescope $A_i$ $(i=1,2,..., n)$ is delayed by the value calculated relative to the Earth center. 
This procedure allows to apply only a geometric delay to the data with an accuracy of one sample time interval.
The delay corresponding to the fractional part of sample time interval is corrected after Fourier transform in the spectral domain.
Next, the phase of signal is shifted consequently to stop interference fringes. This process is called fringe rotation. ASC correlator has two main input parameters: integration time $t_{int}$ and number of Fourier transform channels $N_{FFT}$. $N_{FFT}$ determine the length of data interval in seconds: $T_{FFT}=2 \cdot N_{FFT}\times\eta$, where $\eta=1/(2\Delta f)$ is a Nyquist rate -- a data sampling interval that depends on the $\Delta f$ bandwidth in Hz. The number of FFT data blocks (spectra) for a single integration will be $K=t_{int}/T_{FFT}$.

The sequence of FX correlator operations are the following.

Fringe stopping phase is applied as:
\begin{equation}
\label{a7}
{A_{new}}_i = A_ie^{i\phi_i}, \quad \phi_i= 2\pi\cdot
F_{SKY}\cdot\Bigl(\frac{\tau_{stopFFT}-\tau_{startFFT}}{N_{FFT}}i + \tau_{startFFT}\Bigr),
\end{equation}
where $\tau_{startFFT},\tau_{stopFFT}$ correspond to the delays in the beginning and the end
of given data interval, $F_{SKY}$ is the central observing frequency.

Discrete FFT of each ${A_{new}}_i$ is
\begin{equation}
\label{a8}
F_k = \sum_{n=0}^{N_{FFT}-1}{A_{new}}_n \cdot e^{-\frac{2\pi i}{N_{FFT}}\cdot kn}.
\end{equation}
At this stage the undercompensated part of geometric delay corresponding to the fractional part of time interval sample can be applied to the data.
Fractional bit correction for each ${A_{new}}_i$ sample is
\begin{equation}
\label{a9}
{F_{new}}_i = F_k\cdot e^{\frac{2\pi i\Delta P}{N_{FFT}}},
\end{equation}
where $\Delta P=\frac{\tau}{\Delta \tau}-\Bigl[\frac{\tau}{\Delta
\tau}+0.5\Bigr]$ (squared brackets correspond to the integral part of a number inside them), $\Delta \tau=\tau_{stopFFT}-\tau_{startFFT}$, $\Delta P \subset
[-0.5;+0.5]$.

The next step is to multiply the spectra from two stations:
\begin{equation}
\label{a10}
F_{ij}= {F_{new}}_i\times {F_{new}}_j^\ast,
\end{equation}
where $i, j$ are stations indices, ``$\ast$'' indicates complex conjugation.

As a result, we have cross-correlation spectra determined by number of baselines: $L= n (n-1)/2$. Next, spectrum is averaged as a sum of $K$ spectra:
\begin{equation}
\label{a11_1}
F_{fin}= \frac{1}{K}\sum_{m=0}^{K}F_{ij}^m.
\end{equation}

And the final output will be $L$ average spectra written to disk.

ASC Correlator uses MPI architecture. The principal diagram of interaction between the processes is shown in Fig. \ref {fig1}. Correlator has the ``MAIN'' process that reads the information of a given scan, sends commands to ``READER'' processes. ``READER'' process performs reading of a data for a given telescope for a given scan from the storage and transfer this data to ``WORK'' process. ``WORK'' process receives the data and performs the correlation and ``WRITE'' process writes the correlated data in UVX format to the disk. After correlating all scans ``MAIN'' process sends to all ``READER'' processes a finalization command, which announces the end of work. Next, the process waits for the arrival of ``END'' command and terminates. ASC correlator operate on HPC Cluster with 1 Tflop/s performance.

Arranging of data processing on HPC cluster has the following features:
\begin{enumerate}
\item Raw data is uploaded to high-speed online storage. Then the data is distributed between the cluster nodes. Cluster is capable of running up to 6 parallel correlation jobs;
\item In order to increase write speed the correlated data is recorded on independent hard drives of individual nodes. These drives are configured as RAID 0 (doubles write speed at expense of reliability);
\item Fringe search procedure is executed independently on each node to reduce the load on the internal cluster network.
Thus the structure of the cluster is optimized for maximum splitting of read/write flows to different physical disks since the speed of correlation processing is largely determined by network bandwidth and the data transfer rate than by speed of FFT execution on the computing nodes. 
\end{enumerate}

\begin{figure}[th]
\includegraphics[width=\columnwidth]{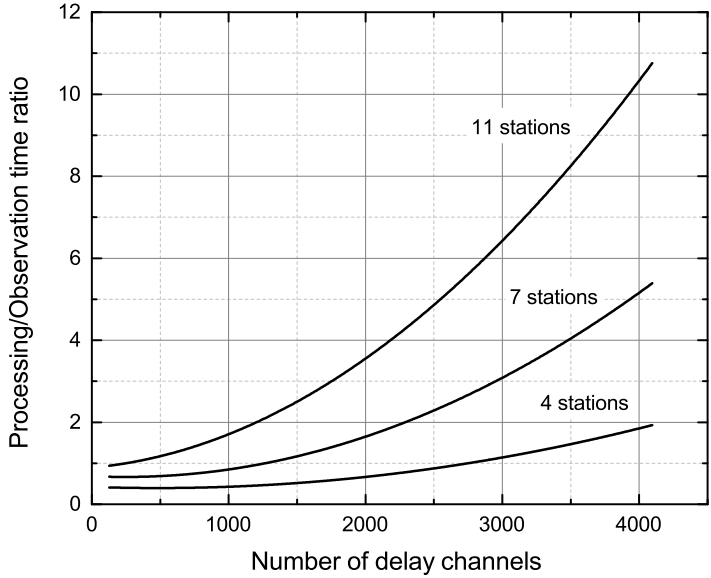}
\caption{\small ASC correlator performance expressed in ratio between processing time and duration of observation. Three curves show how performance depend on the number of stations and number of delay channels.} \label{fig2}
\end{figure}

ASC correlator performance is determined by number of stations in observation, number of Fourier transform channels and integration time. Correlation and fringe search procedures for Radioastron data are strongly connected to each other. The first run of correlator is performed in broader window of delay and fringe rate in order to search for the fringe. Large number of participating ground telescopes generates a big input data rate. Therefore it is necessary to pay attention to the power of disk subsystem and network bandwidth between the cluster nodes and cluster to raw data storage. A particular challenge is the limited capacity of raw data storage and the data transfer rate from offline hard drives to online storage. Fig. \ref{fig2} shows correlating efficiency depending on the number of telescopes in observations. This figure shows the load only on cluster nodes. The current performance is sufficient to meet all the needs of the mission.

ASC correlator has the following properties:
\begin{enumerate}
\item Full support of widely distributed VLBI data formats: RDF (Radioastron data format), Mark5A, Mark5B, VDIF, VLBA, K5;
\item Cluster with 1 Tflop/s performance;
\item Online raw data storage of 80 TB for correlation. Baseband/voltage data is uploaded to this storage from offline raw data archive in order to perform the correlation.
\item Online data storage for correlated observations -- 160 TB;
\item Raw data archive of 2450 TB (hard drives) and 2000 TB (tapes);
\item Real time data processing of up to 10 stations (45 baselines, 256 Mb/s bitrate) with delay window of 256
channels. Typical session has 4 stations and 2048 Fourier transform channels. Processing rate in this case is two times faster than real time.
\item Correlator can operate in the following modes:
\begin{enumerate}
\item Continuum.
This mode is used to correlate the observational data of compact Galactic and extragalactic radio sources (active galactic nuclei, quasars).

\item Spectral line.
This mode is used to correlate the data of maser observations, which require higher spectral resolution to achieve the required velocity resolution of 0.1 -- 0.3 km/s. Data is processed in two passes. The first pass is the same as for continuum mode. The second pass is performed in a selected frequency range that contains spectral details in order to search for fringes from spectral lines. The chosen spectral resolution is determined by the number of Fourier transform channels and depends on the observing frequency and properties of the specific spectral line source.

\item Pulsars.
Pulsars are emitting in a very short time and a correlation window can be used to increase the signal to noise ratio. In this case correlation is performed in so-called ``Gate mode''.

Features of pulsar data processing are related to the dispersion effect caused by free electrons located on the line of sight. Dispersion lead to the delay of signal in frequency. There are two methods to compensate this effect: coherent and incoherent dedispersion.

While coherent dedispersion is applied to the whole observed frequency band, incoherent dedispersion divides frequency band into a number of small channels (bins). Signal in each channel is shifted in time to compensate for the difference in pulse time of arrival, after that the channels (bins) are summed. The final time resolution in this case is limited by the size of a single bin.

Incoherent dedispersion in ASC correlator is performed in several steps. Pulsar period is divided into channels (bins). Correlator performs the calculation of spectra for each bin. The delay of signal in each bin is compensated by polynomial model calculated with TEMPO2 software package \cite{Hobbs,Edwards}. Then the signal in each bin is being averaged over the observation time. As a result, each bin will contain the signal with removed dispersion. Dispersion effect will decrease linearly with increasing number of bins.

Accordingly it is necessary to sum only the data that contain fringes of pulses and discard the rest. It is important to compensate the pulse shape with a dispersive delay.

ASC correlator has number of different pulsar data processing modes:
\begin{enumerate}
\item  Simple Gate Mode: 
Single gate is set on pulse. Increase in S/N ratio can be up to 3-5 times;
\item  Compound Gate Mode: 
Gate is weighted according to the shape of pulsar average profile. This approach can increase S/N up to 6-20\% over simple gate mode;
\item   Bins Mode:
This mode use many gates (bins). In case the pulse phase is unknown this mode can be used to determine the mean profile of the pulsar. When the pulse phase is known this mode is useful for simultaneous extraction of on pulse and off pulse data. Off pulse data can be treated as normal continuum data and used for bandpass calibration.
\item    Giant Pulses Mode:
This mode is similar to bins mode, but after the correlation in each bin the correlator compares correlation coefficient with a certain threshold, and save the data only for those moments in time and those bins, which have correlation coefficients exceeded the given threshold.
\end{enumerate}

\item ``Coherent'' mode. This mode can be used as an alternative way to synchronize an onboard reference frequency using the H-maser of ground tracking station. This mode is important, because it is used as a backup in case of an unexpected failure of onboard H-maser. Main stages of ``Coherent'' mode are shown on Fig. \ref{fig3}:
 
\begin{enumerate}
\item Reference signal is transmitted from ground based H-maser to the space radio telescope: $f_{up} = f_0+\Delta f_c$ , where  $f_0 =7.2075$ GHz, $\Delta f_c$ is Doppler correction term calculated each 40 ms from the space radio telescope orbital data for each transmission moment  $t''$.
\item At instance $t'$ is delayed by $\tau_u$, signal at $f_u = 7.2\,\textrm{GHz} +\Delta f_c + \Delta f'_d$ is received by the space radio telescope, where $\Delta f'_d$ is Doppler frequency shift on the route from the tracking station to the space radio telescope.This signal is multiplied by a factor of $(1+q_T)$ and coherently converted to 8.4 GHz frequency band. The coefficient $q_T$ is equal to $159/961$.
\item The signal $f_b = (1+q_T)\cdot f_u$ is transmitted down from the space radio telescope to the tracking station at 8.4 GHz band and finally received by tracking station at moment $t$ with delay $\tau_D$ relative to the moment $t'$.
\end{enumerate}

Received frequency will be $f_r = f_b + \Delta f_d$, where $\Delta f_d$ is Doppler frequency shift on the route from the space radio telescope to the tracking station at the moment $t$. Errors of predicted orbit result in residual Doppler frequency shift between transmitted and received signals $f_{RES} = f_{up} - f_r$.

Residual phase can be obtained from residual frequencies by integration on time interval $(t - t_0)$:
\begin{equation}
\label{a11}
\Delta\varphi_{RES}=\int_{t_0}^t f_{RES}(x)dx
\end{equation}
where $t_0$ is the beginning of observations.

Consequently additional shift of the tracking station master clock can be calculated for the moment $t_i$ using expression: $x(t) = \Delta\varphi_{RES}(t)/(2\pi f_0)$. The corrected delay $D(t_i)$ at the moment $t_i$ can be estimated as  $D(t_i) =d(t_i) + x(t)$, where $d(t_i)$ -- is a delay calculated from reconstructed orbit. Two examples of delay recovery are presented in Fig. \ref{fig4} and Fig. \ref{fig5}. Both figures demonstrate the results of ``Coherent mode'' applied to observations conducted with Radioastron at C- and K- bands correspondingly.

\begin{figure} [th]
\includegraphics[width= \columnwidth]{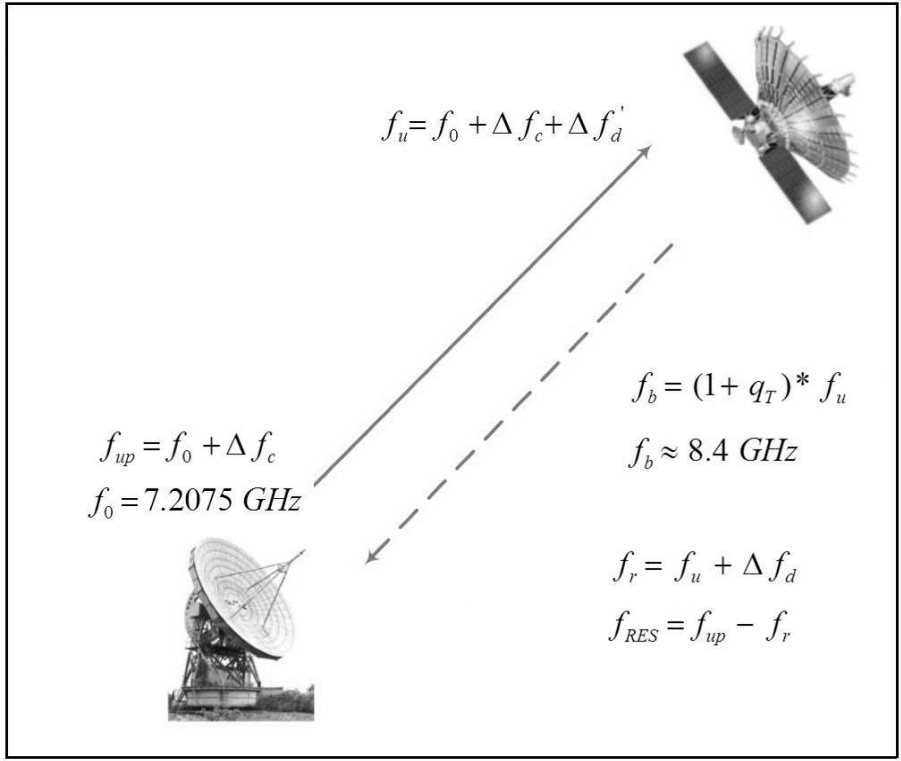}
\caption{\small Scheme of ``Coherent'' mode implemented in ``Radioastron" project.} \label{fig3}
\end{figure}

\begin{figure}[th]
\includegraphics[width= \columnwidth]{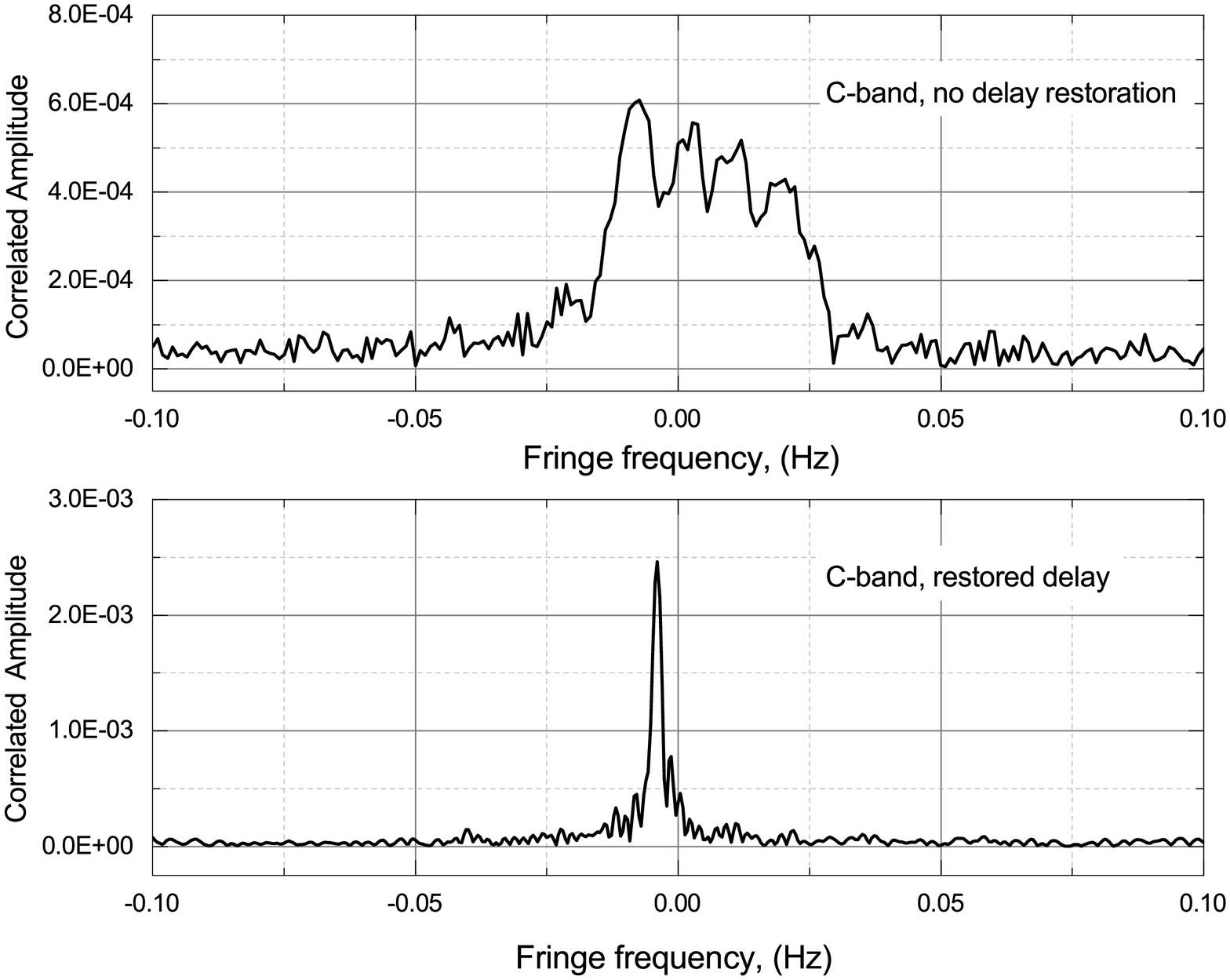}
\caption{\small Example of fringe for Radioastron ``Coherent'' mode. Observation code: RAES02AD. Source 2013+370. Date: 12.05.2012 07:00 - 08:30. Baseline Radioastron-Effelsberg, baseline projection 0.6xED. Observation wavelength $\lambda$ = 6 cm (C-band).  Top - no time correction applied. Bottom - time correction applied.} \label{fig4}
\end{figure}

\begin{figure} [th]
\includegraphics[width=\columnwidth]{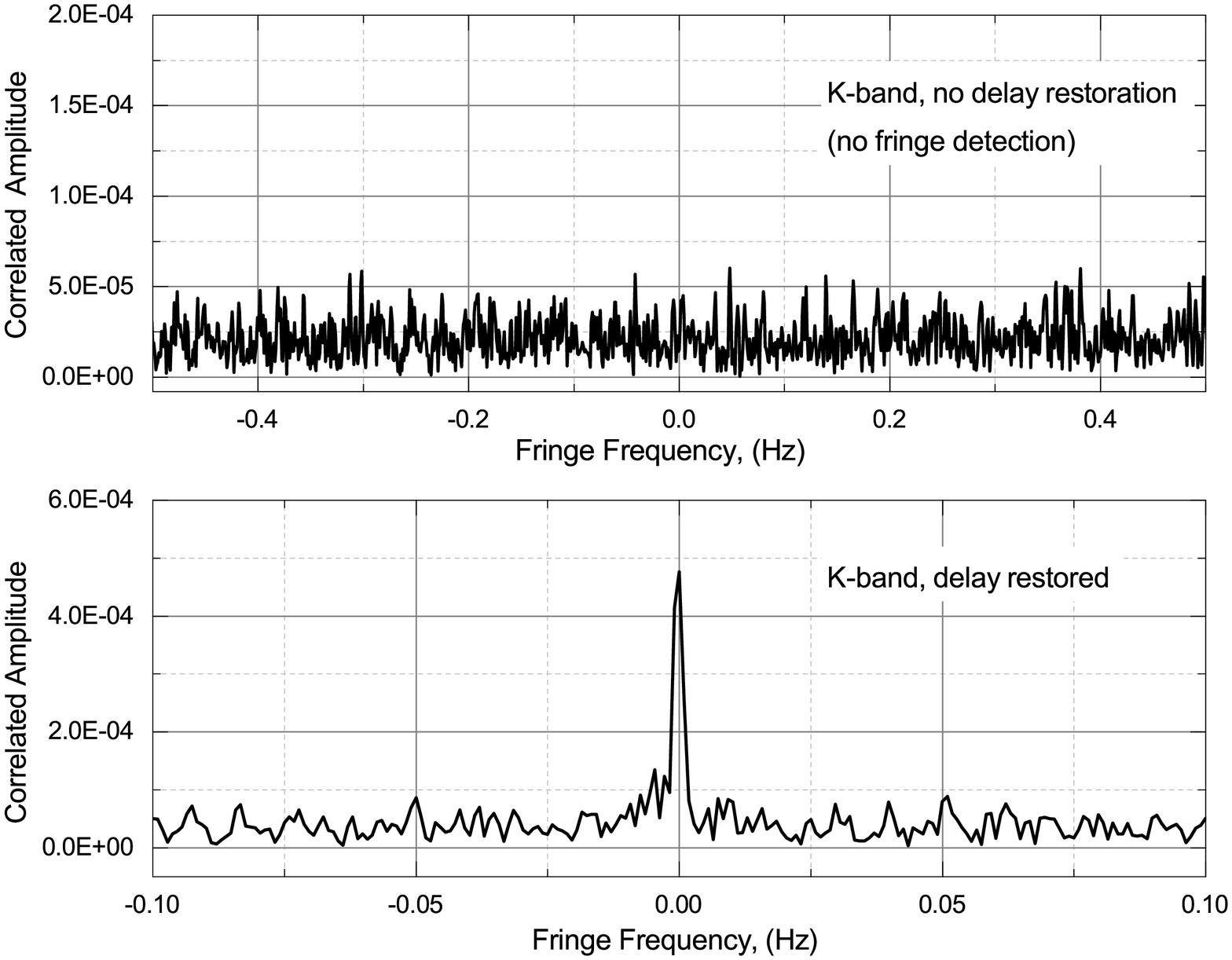}
\caption{\small Example of fringe for Radioastron ``Coherent'' mode. Observation code: RAES02AD. Source 2013+370. Date: 12.05.2012 07:00 - 08:30. Baseline Radioastron-Effelsberg, baseline projection 0.6xED. Observation wavelength $\lambda$ = 1.35 cm (K-band). Source 2013+370. Top - no time correction applied. Bottom - time correction applied.}
\label{fig5}
\end{figure}

\end{enumerate}
\end{enumerate}

\subsection{Correlation in practice. Delay estimation.}

Correlation results rely on the delay model. For space-ground VLBI the accuracy of interferometer delay model depends directly on the quality of reconstructed orbit. Inaccuracies of delay calculation are the matter of errors in position, velocity and acceleration of the SRT. Orbit quality imposes significant constraints on the success of fringe-delay search. Correlator performs the correlation in a given range of delay and fringe rate values, which is determined by the integration time and the number of Fourier transform channels. Selection of integration time and number of Fourier transform channels is directly associated with the errors in the space radio telescope state vector. State vector is a cartesian coordinate system including the position $\overrightarrow{r}$ and velocity $\overrightarrow{v}$ that together with time (epoch,  t) determine the trajectory of the orbiting body in space.
The maximum size of correlation window is only limited by computer performance. Such feature allows to get almost stable fringes in the presence of errors in the space radio telescope state-vector. 

The delay window of the correlator depends on the size of FFT and is given by $W_{del} = N_{FFT} / (2 \cdot \Delta f)$. The fringe rate window of the correlator depends on the integration time $t_{int}$ and is given by $W_{frr} = 1/2 \cdot t_{int}$.

Error in position of the space radio telescope $\Delta{\overrightarrow{x}}_s$ gives shift in delay
$\tau_{err}$
\begin{equation}
\label{a14}
\tau_{err} = {\overrightarrow{K}}\cdot  \Delta{\overrightarrow{x}}_s / c.
\end{equation}
where $\overrightarrow{K}$ -- the space radio telescope position vector and $\Delta {\overrightarrow{x}}_{s}$ - error of position vector.

\begin{figure}[th]
\includegraphics[width=\columnwidth]{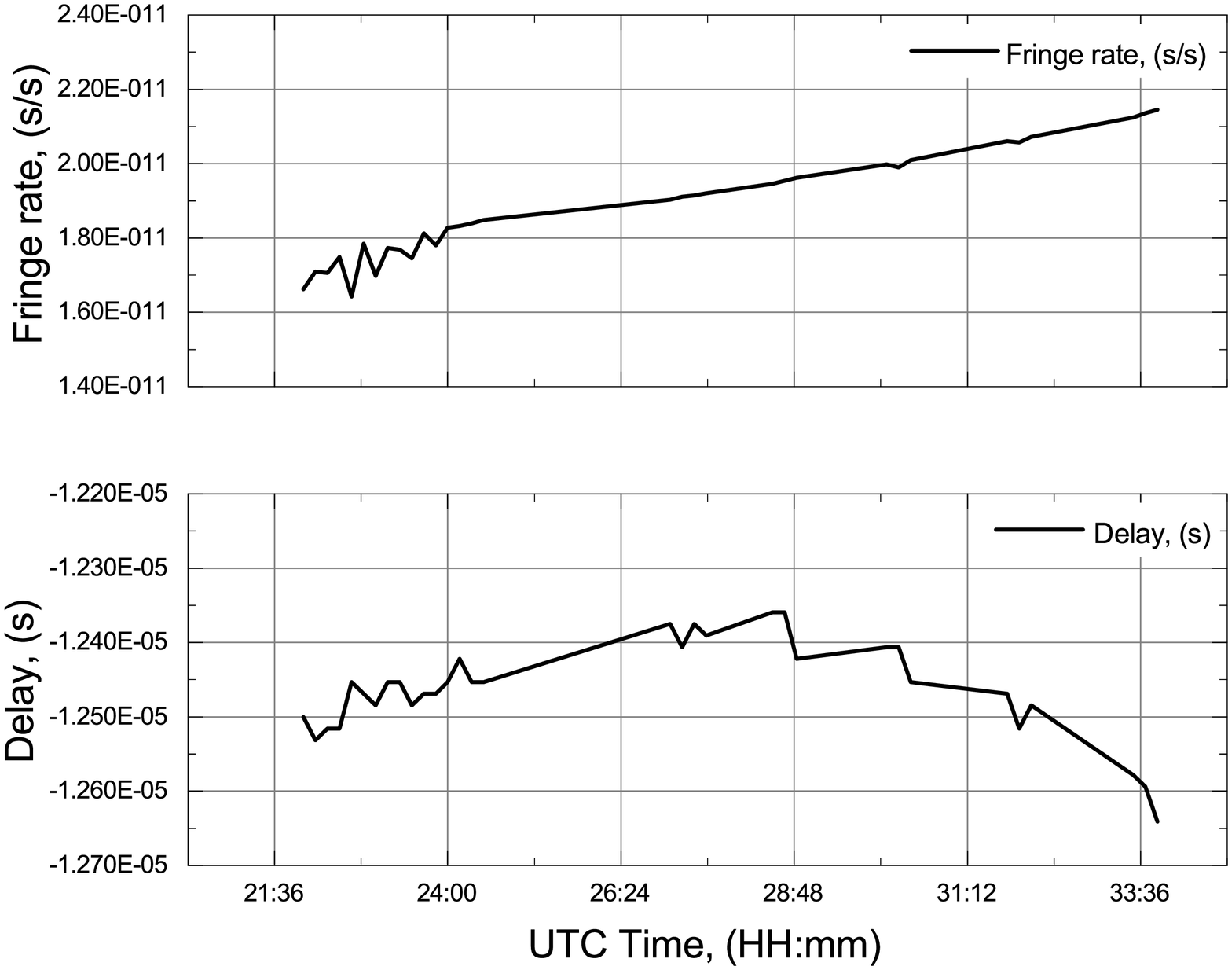}
\caption{\small Example of observed errors in delay and delay rate. Observation code: RAKS11AA. Source: 0716+714. Date: 03.01.2015 22:00 - 04.01.2015 10:00. Observation wavelength $\lambda = 1.35$ cm (K-band). Trend of delay $\pm 10^{-7}$ s  (bottom) and delay rate $4\cdot 10^{-12}$ s/s (top) correspond to $\pm 3$ m error of the spacecraft position and $\pm 12$ mm/s error in velocity.} 
\label{fig7}
\end{figure}

An example of observed delay error of $\pm10^{-7}$ s and fringe rate error of $4\cdot10^{-12}$ s/s is shown on Fig. \ref{fig7}. It is seen that those errors of delay and fringe rate are equivalent to $\approx 3$ m error in position and $\approx$ 12 mm/s error in velocity.

The SRT orbit uncertainties of 2 cm/s in velocity and 200 m in position determine the choice of correlation parameters for K-, C-, L- and P- bands (Table \ref{tab3}):
\begin{table}[th]
\center
\caption{Correlation parameters in  K, C, L and P bands}\label{tab3}    
\begin{tabular}{|l|c|c|c|c|c|}
\hline
Band & FFT      &Delay           &Integration  &Fr. Rate    &Data volume, \\
     & channels &window ($\mu$s) &time (s)   &window (Hz) &1 h observation (Gb) \\
\hline
K & 2048 &64 &1/64 &32  &420\\
C & 2048 &64 &1/8  &4   &52\\
L & 2048 &64 &1/4  &2   &26\\
P & 2048 &64 &1    &0.5 &6.6\\
\hline
\end{tabular}
\end{table}

\begin{figure} [th]
\includegraphics[width=\columnwidth]{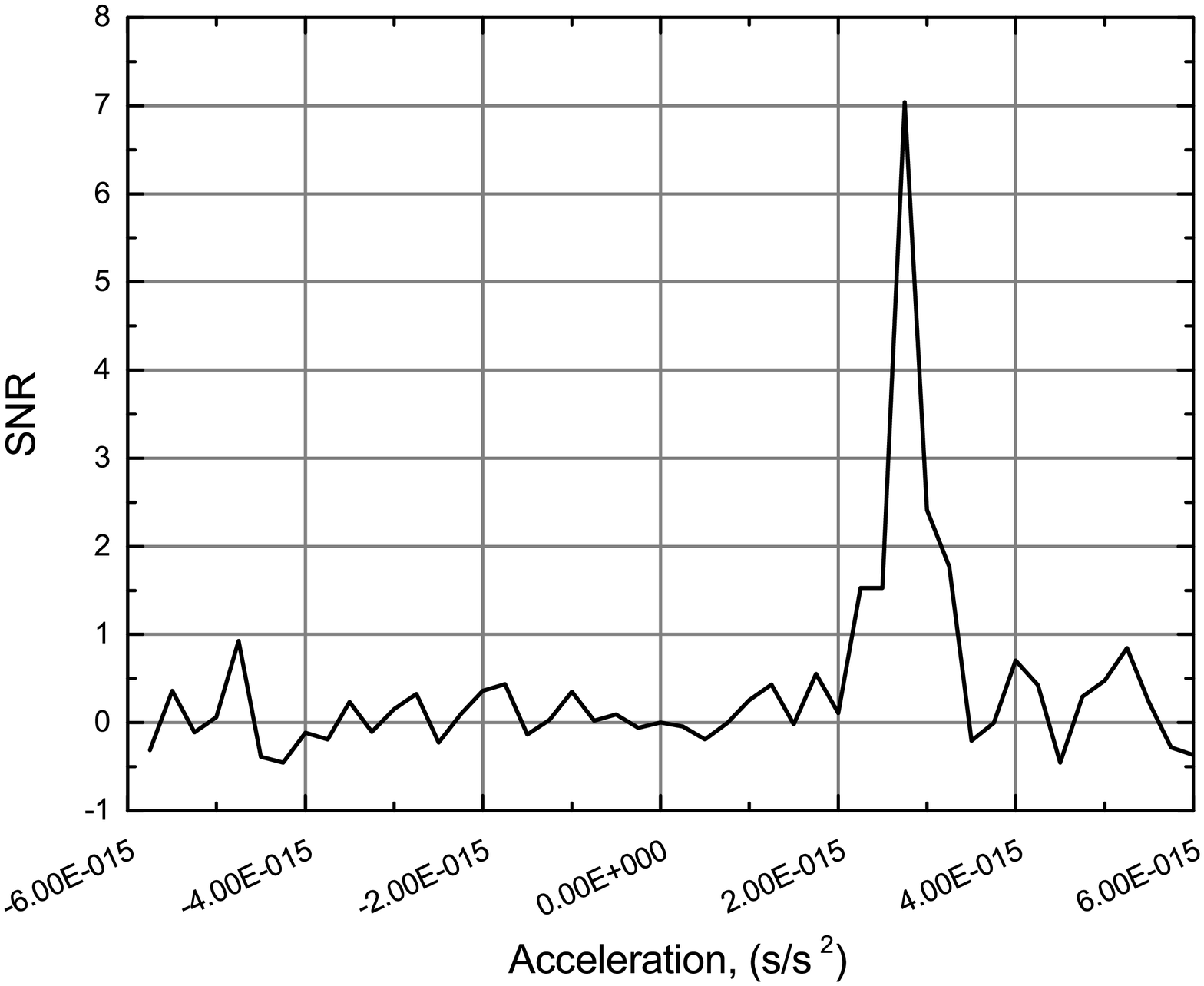}
\caption{\small Estimation of the acceleration. Peak SNR corresponds to the best found value of acceleration for one single scan (570 s).}
\label{fig8}
\end{figure}

Correlation procedure requires to take into account uncertainties related to the space radio telescope acceleration. Search for the acceleation in correlation window can significantly increase the signal-to-noise ratio (Fig. \ref{fig8}). The correlator has a procedure that sequentially step-by-step calculate the acceleration in correlation window in order to find fringe with the highest signal-to-noise ratio.

Radioastron experience showed that it is necessary to perform the first correlation pass with much wider correlation window, than it is used for ground baselines. This is caused by significant uncertainties of orbit determination. The space radio telescope orbit uncertainties results to the increasing size of correlation window that leads to higher computing load but does not cause irretrievable losses of scientific result. Another way to avoid fringe loss caused by orbit errors is to perform simultaneous observations at two different frequencies. This approach become very frequent in Radioastron mission. In case correlation is detected at longer wavelength, the corrections to the delay model can be used for the correlation procedure of shorter wavelength increasing the chance of fringe detection. Correlated observations are recorded to the output data archive of Radioastron mission as UVX and IDI-FITS files for further astrophysical interpretations using various post-correlation data analysis software applications (e. g. ASL, AIPS, PIMA etc.).

\section{Delay calculation for space-ground VLBI}\label{Reduction}

All correlators require a precise method for calculation of the geometric delay which is then applied to align the data streams from the different telescopes  or for delay tracking and fringe stopping for correlation of the signals. For astronomical VLBI most correlators use CALC software model to calculate the delay for these purposes~\cite{Gordon}.

ASC Correlator use ORBITA2012 delay model based on ARIADNA algorithms. The main differences are related to the computation of delay for space-ground baselines (see Chapter 4 in ~\cite{Zharov}).

ARIADNA performs calculation of delays between the telescopes (baseline-based delay calculation). Calculation of baseline-based delay is performed with ``consensus model'' (see Chapter 11 in \cite{IERS2010}). ``Consensus model" defines the standard reference systems implemented by the International Earth Rotation and Reference Systems Service (IERS) and the models and procedures used for this purpose. The reference systems and procedures of the IERS are based on the resolutions of International Scientific Unions \cite{IERS2010}.
ORBITA2012 calculates the delays between individual telescopes, including the space radio telescope (antenna-based delay) related to the Earth center. In this case it is easy to calculate a baseline delay if two antenna-based delays are known. Situation with delay calculation for the SRT is more complicated, because of the uncertanities in the predicted orbit of the SRT and presence of drift related to the frequency variations of the onboard H-maser. 

The antenna-based delay model is represented as a polynomial function in time with coefficients:

\begin{equation}
\label{a15}
\tau(t) = a_0+a_1t+a_2t^2+\ldots +a_nt^n.
\end{equation}

Polynomial order $n$ can be adjusted (default is $n=6$) and coefficients $a_0, a_1, \ldots a_n$ are calculated every period, where t=0 is defined as a start of each period. By default this period is equal to 1 minute and can be adjusted by the operator.

Conventional displacements of ground telescopes in ORBITA2012 delay model include all known correction terms, such as: tectonic and tidal motions (mostly near diurnal and semidiurnal frequencies), non-tidal motions associated with displacement due to ocean tidal loading and due to diurnal and semidiurnal atmospheric pressure loading, etc.~\cite{IERS2010}.

Spacecraft state-vector is calculated by KIAM with an accuracy of $\sim 200$ m in position and $\sim 2$ cm/s in velocity for the final (reconstructed) orbit~\cite{Duev}. Accuracy of the spacecraft state-vector is critical for successful correlation of Radioastron observations. Development of consistent dynamical model of the space radio telescope is associated with a number of challenges. The most important challenge is defined by a significant amount of pressure on the spacecraft surface caused by solar radiation, which produces both perturbations of acceleration and torque. They are compensated by the stabilization system that maintains constant attitude with respect to an inertial frame using reaction wheels. Unloadings of reaction wheels generally occur one or two times per day by firing thrusters. This leads to orbit perturbations. Change in velocity of the spacecraft's center of mass is about $3-5$ mm/s in magnitude during unloading. Both perturbations caused by solar radiation and unloadings of reaction wheels can lead to a significant orbital inaccuracy.

Data from ground tracking stations and data from onboard measurements are used to improve the initial state vector (position and velocity) of the space radio telescope. Routine tracking is performed by two Russian stations located near Ussuriysk (70-m antenna) and in Bear Lakes (64-m antenna). The data usually contain two-way ranging and Doppler measurements. In addition tracking stations located in Pushchino, Russia and in Green Bank, USA can be used to collect the Doppler measurements.

\section{Comparison between ASC and DiFX correlator}
\label{comparing}
Comparison of ASC correlator with DiFX was performed at the final stage of correlator development and during the first cycle of Radioastron mission operation. 

\begin{figure}[th]
\center
\includegraphics[width=\columnwidth]{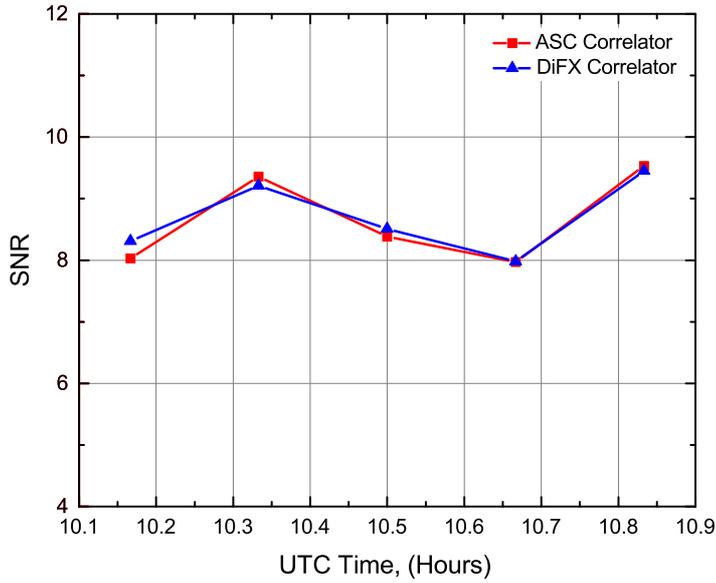}
\caption{\small Comparison of SNR between ASC and DiFX correlators. C-band, Baseline: Radioastron-Arecibo. Relative difference is not more than $\approx$ 2\% during time interval of $\approx$ 40 min.} \label{fig9}
\end{figure}

Results of comparison showed, that after the second correlation pass the difference of signal-to-noise ratio between ASC and DiFX outputs does not exceed 1.5 - 2\% (see Fig. \ref{fig9}).

\begin{figure} [th]
\includegraphics[width=\columnwidth]{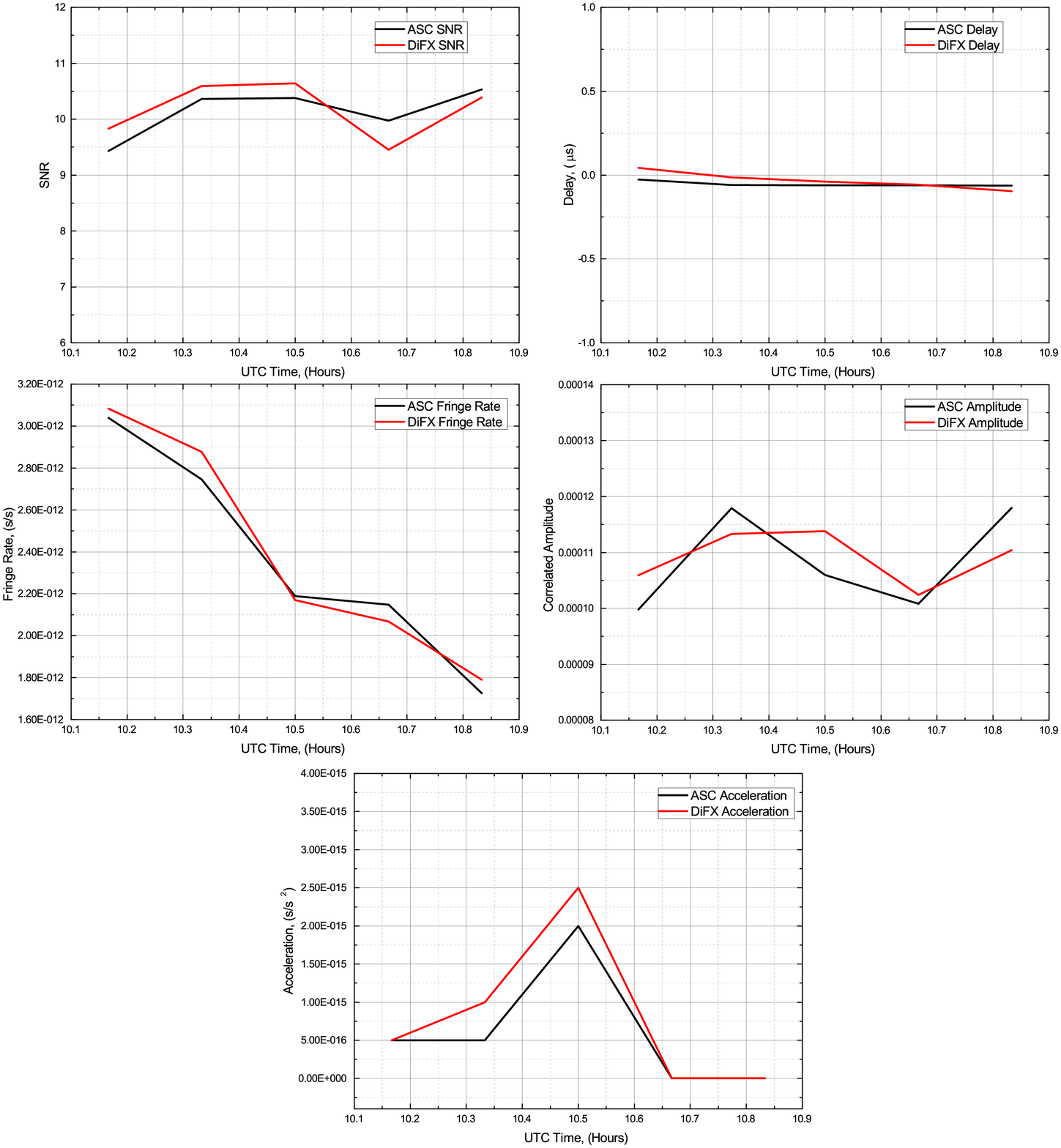}
\caption{\small Example of comparison between ASC and DifX correlators output for space-ground baseline. Observations code: RAES03EV. Source: 0748+126. Date: 24.10.2012 09:00 - 10:00. Observing wavelength: $\lambda=6$ cm. Baseline: Radioastron-Arecibo. Baseline projection: 11xED. The following values of ASC and DiFX correlators were compared: SNR (left, top), delay (right, top), fringe rate (left, middle), correlated amplitude (right, middle), acceleration (bottom).}
\label{fig10}
\end{figure}

\begin{figure} [th]
\includegraphics[width=\columnwidth]{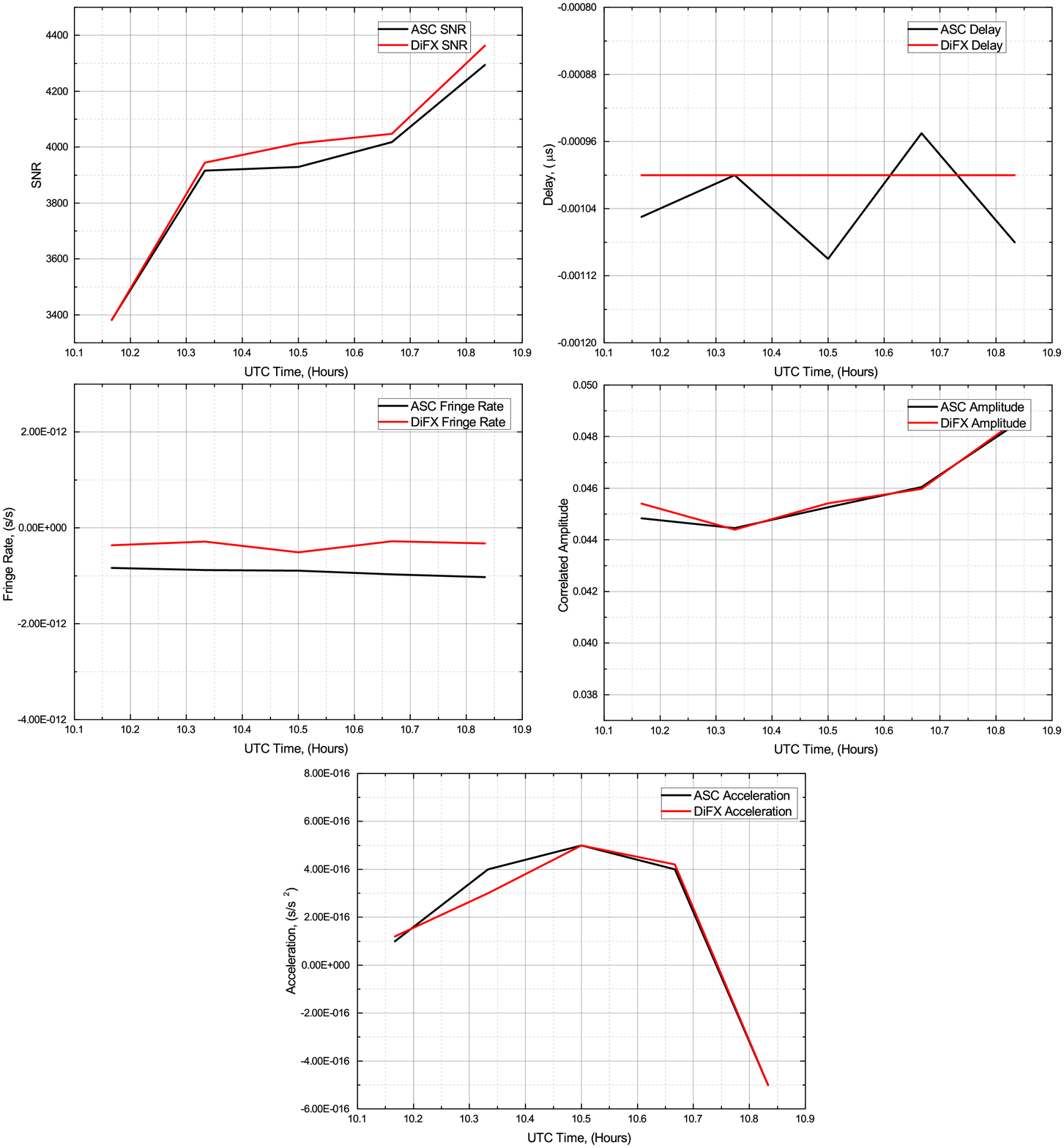}
\caption{\small Example of comparison between ASC and DifX correlators output for ground baseline. Observations code: RAES03EV. Source: 0748+126. Date: 24.10.2012 09:00 - 10:00. Observing wavelength: $\lambda=6$ cm. Baseline: Yebes-Arecibo. The following values of ASC and DiFX correlators were compared: SNR (left, top), delay (right, top), fringe rate (left, middle), correlated amplitude (right, middle), acceleration (bottom).}
\label{fig11}
\end{figure}

In our analysis, we compared a number of parameters that mainly characterize the quality of correlator output data: correlated amplitude, signal-to-noise ratio, residual delay, residual delay rate (fringe rate) and residual acceleration (delay 2-nd derivative).
Example of such comparison is shown in Fig. \ref{fig10} and Fig. \ref{fig11} for space-ground and ground baselines correspondingly.

\begin{figure} [th]
\includegraphics[width=\columnwidth]{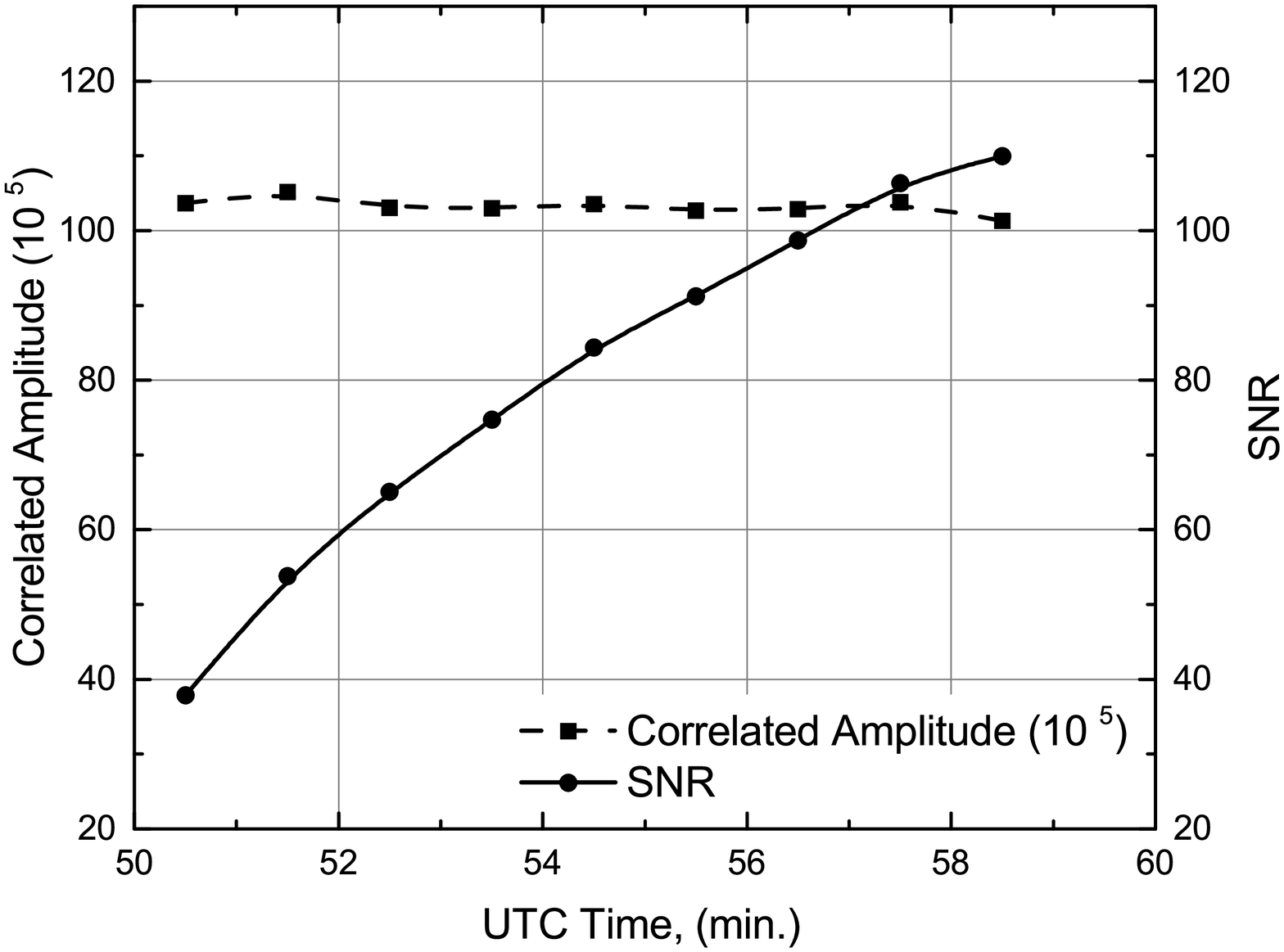}
\caption{\small 
Behavior of correlated amplitude and SNR during integration on a 10 min scan. Observation code: RAES03FU. Source: 0716+714. Date: 19.11.2012 00:20 - 01:00. Baseline Radioastron-Svetloe, baseline projection: 2.5xED. Observing wavelength: $\lambda=6$ cm. Amplitude has stable behavior during 10 min coherent integration. SNR shows dependence close to the theoretical integration law $1/\sqrt {t}$.}
\label{fig12}
\end{figure}

One of the important features of VLBI method is its sufficient phase stability that enables long integration of the correlation function. Signal that is received by Radioastron antenna located in open space isn't affected by atmospheric distortions. Such advantage, as well as a stable onboard H-maser provides longer integration times, which is quite important for higher frequencies like 22 GHz (see Fig. \ref{fig12}).

\begin{figure} [th]
\includegraphics[width=\columnwidth]{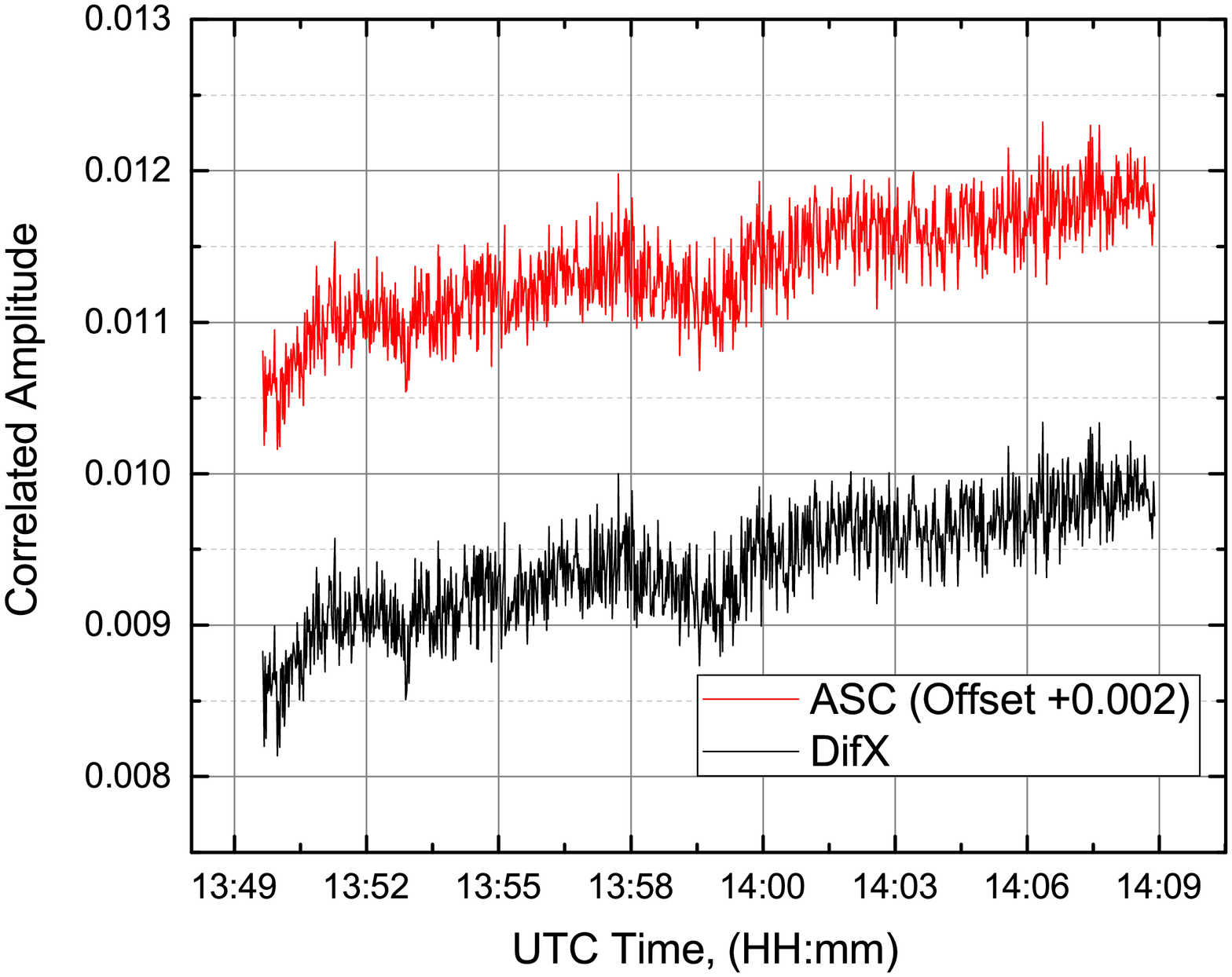}
\caption{\small Comparison of visibility amplitude between ASC and DiFX correlators. Observation wavelength $\lambda = 18$ cm (L-band). Baseline Westerbork-Effelsberg. Deviation does not exceed 1.5 \% on time interval of $\approx$ 1200 s. ASC correlator amplitude (red) has an offset of +0.002 for easy viewing.}
\label{fig13}
\end{figure}

Additional comparison of ASC and DiFX visibilities showed adequate correspondence of signal-to-noise ratio, as well as a good correspondence of visibility amplitude behavior in time (see Fig. \ref{fig13}).

More details on comparison of ASC, JIVE SFXC and DiFX correlators will be given in a separate publication.

\section{Discussion}~\label{Concl}

During 2011 - 2016 years under active regulations of Russian Federal Space Agency, Lavochkin Association and many participating scientific international organizations number of Radioastron observing programs were successfully carried out:
\begin{enumerate}
\item Fringe Search program (sessions code RAFS), 11.2011 -- 02.2012;
\item Early Science program (sessions code RAES), 02.2012 -- 06.2013;
\item Radioastron Key Science Program (sessions code RAKS), executed from 07.2013;
\item Radioastron General Observing Time (sessions code RAGS), executed from 07.2014.
\end{enumerate}

\begin{figure} [th]
\includegraphics[width=\columnwidth]{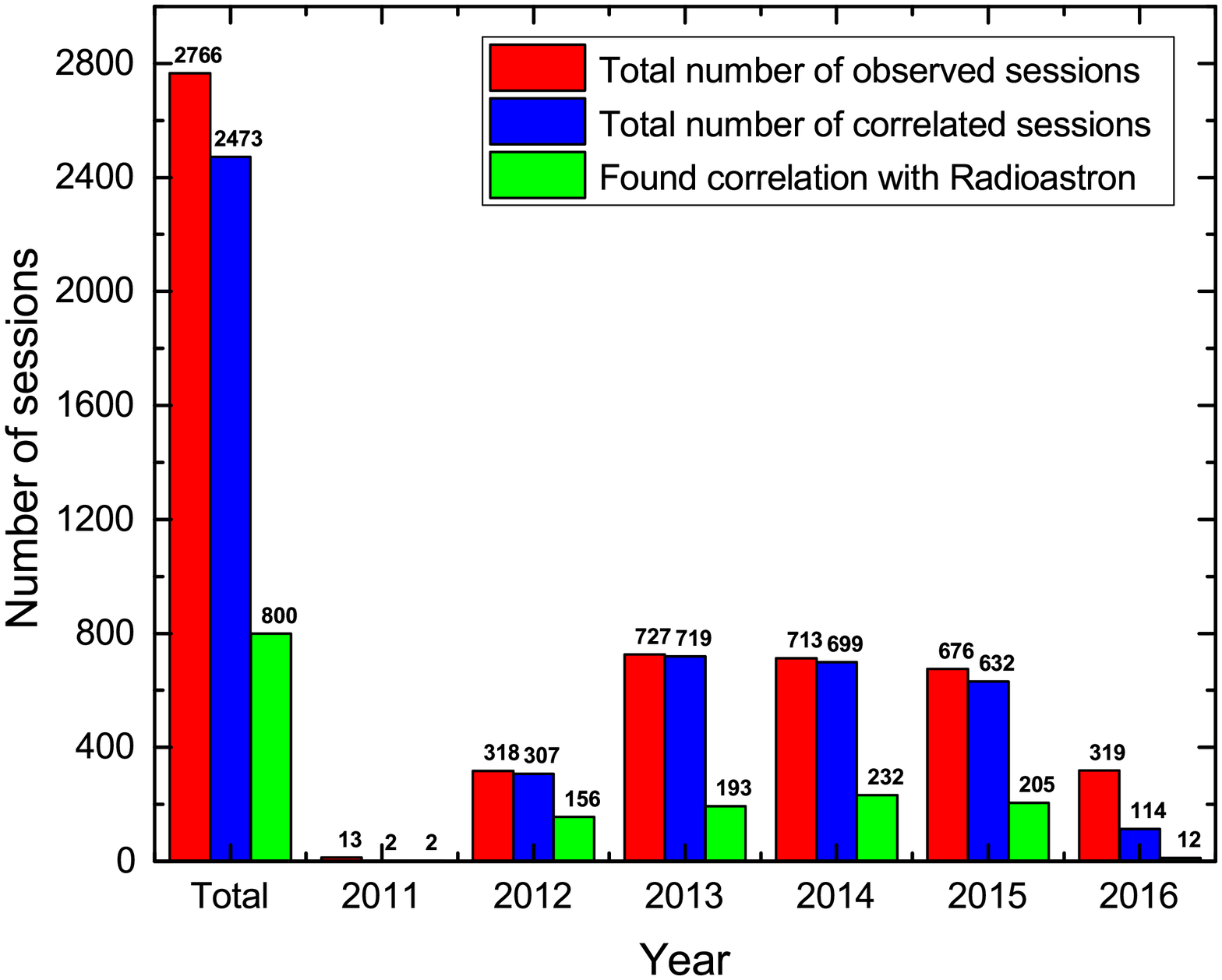}
\caption{\small Number of yearly correlated Radioastron space-ground observations using ASC correlator.
Covered period: November, 2011 -- May, 2016. About 30\% of all correlated observations yield fringe detections with the space radio telescope.}
\label{fig14}
\end{figure}

\begin{figure} [th]
\includegraphics[width=\columnwidth]{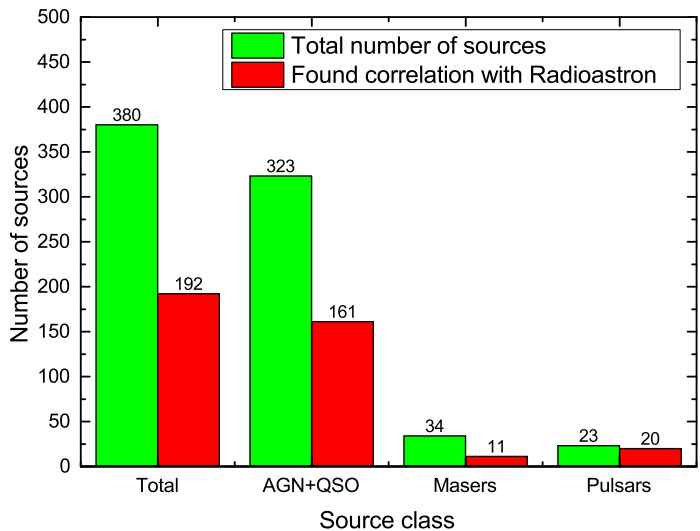}
\caption{\small Number of unique source observed with Radioastron space-ground VLBI. Represented period: November, 2011 - May, 2016.
About 50\% of unique source yield fringe detections with the space radio telescope.}
\label{fig15}
\end{figure}

Fig. \ref{fig14} show a total amount of observations, including sessions with successful fringe detections with the SRT. Number of observed unique sources is shown in Fig. \ref{fig15}. About one half of AGN+QSO sources provide statistically reliable correlation results.

\begin{figure} [th]
\includegraphics[width=\columnwidth]{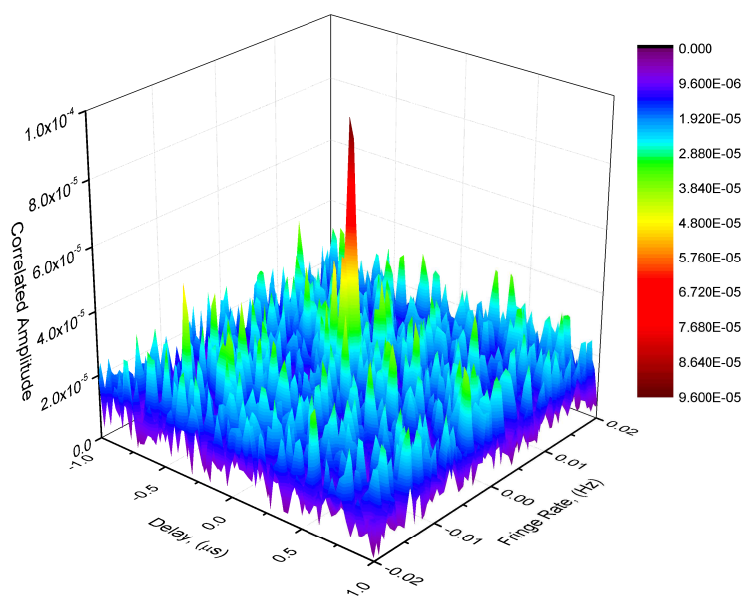}
\caption{\small Delay-Fringe rate diagram for blazar 3C279. Distance to this source is 3070 Mpc.
Observation code: RAES11A. Date: 02.02.2013 07:10 - 08:40. Baseline Radioastron-Effelsberg, baseline projection 9.4xED. Observation wavelength $\lambda=1.35$ cm. Angular resolution 18 $\mu as$. Estimated source linear size 0.11 pc.}
\label{fig16}
\end{figure}

\begin{figure} [th]
\includegraphics[width=\columnwidth]{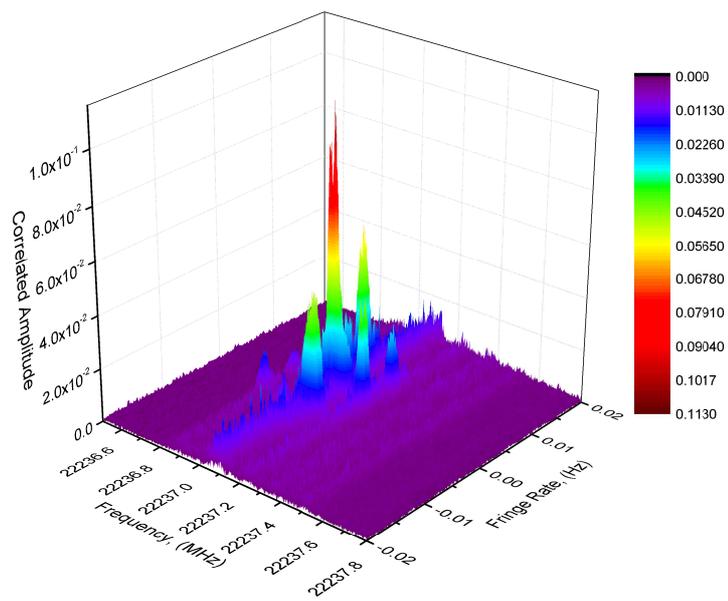}
\caption{\small Frequency-Fringe rate diagram for galactic maser W3IRS5. Observation code: RAES02AB. Date: 30.12.2012 16:40 - 17:20. Baseline Radioastron-Effelsberg, baseline projection 4xED. Observation wavelength $\lambda=1.35$ cm. Angular resolution 40 $\mu$as. Estimated maser source linear size $1.2\cdot10^7$ km.}
\label{fig17}
\end{figure}

\begin{figure} [th]
\includegraphics[width=\columnwidth]{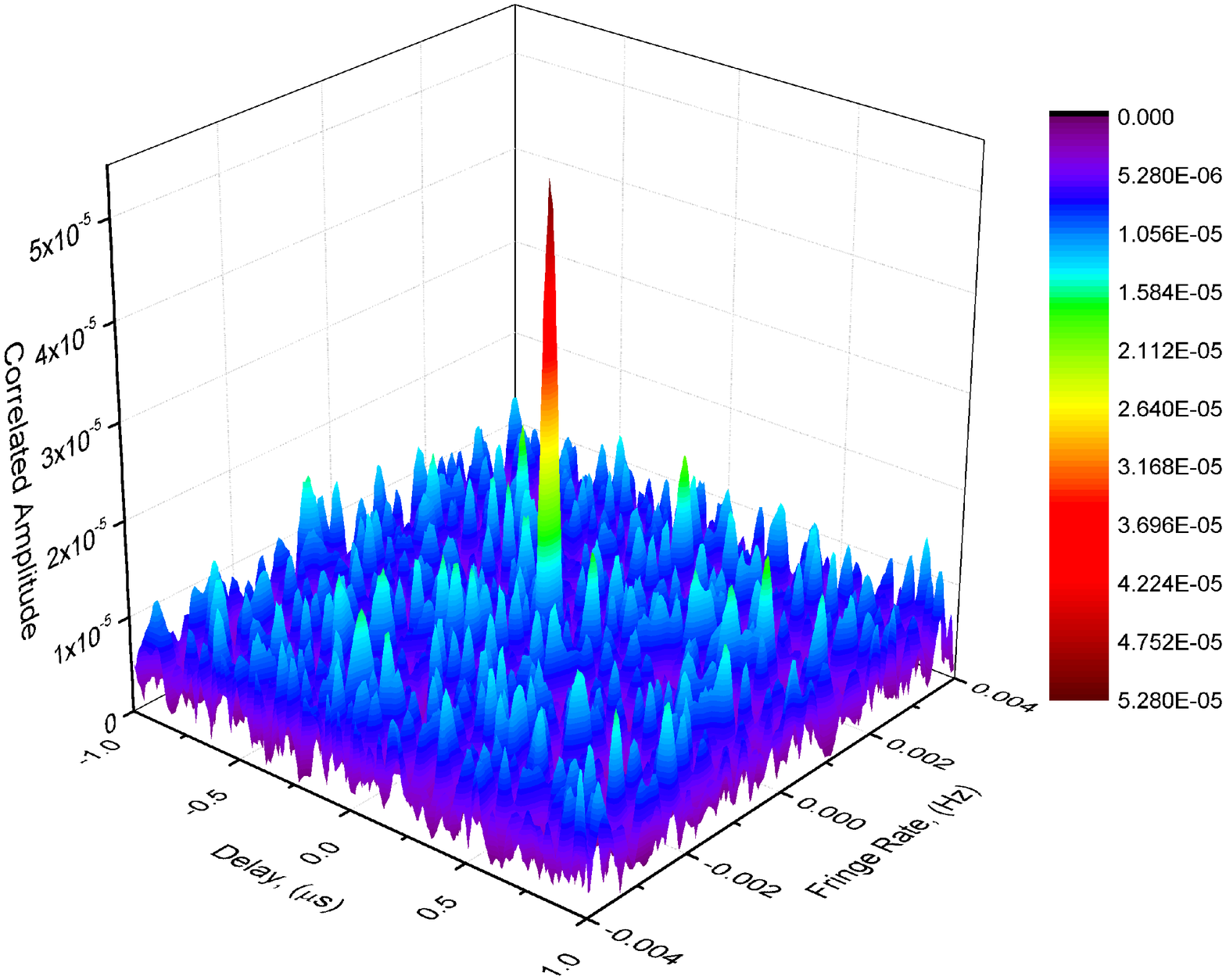}
\caption{\small Delay-Fringe rate diagram for 0048-097. Observation code: RAKS01KT. Date: 01.12.2013 21:00 - 22:00. Baseline Radioastron-Green Bank, baseline projection 26.5xED. Observation wavelength $\lambda=18$ cm.}
\label{fig18}
\end{figure}

Fig. \ref{fig16} and \ref{fig17} show 3D diagrams of fringes for blazar 3C279 (baseline projection 9.4 Earth diameters) and Galactic maser W3IRS5 (baseline projection 5.28 Earth diameters) correspondingly. Fig. \ref{fig18} shows fringe detected on Radioastron-Green Bank baseline of 26.5xED at $\lambda=18$ cm wavelength.

Comparison between ASC and DiFX correlators showed good compliance of correlation results. The deviations does not exceed 1.5\%.

Delay calculation algorithm in ASC Correlator is flexible and can be applied for different correlation modes: continuum, spectral line, pulsar, ``Coherent'' mode. Outstanding angular resolution was achieved -- the interferometer fringes were obtained at extreme baseline projections at K-band ($\sim$14 $\mu$as):
\begin{enumerate}
\item 345 000 km (27.1xED) at L-band,
\item 295 000 km (23.1xED) at C-band,
\item 197 000 km (15.5xED) at K-band.
\end{enumerate}
Till the 1st of May 2016 2766 sessions were successfully correlated. Fringes at space-ground baseline were found for more than 800 observations ($\sim$ 30\%).

The maximum data processing rate in Radioastron mission is about 250 sessions per month, while the observation rate is about 80 - 120 experiments per month. Thus ASC Correlator performance is enough to correlate all the data of Radioastron project without significant delays. Space-ground VLBI data correlation in contrast to ground VLBI  largely depends on network performance and data delivery, rather than on the computational performance.

\section{Conclusions}
Astro Space Center (ASC) software FX correlator is an important component of space-ground interferometer for Radioastron project. It was developed in Astro Space Center of P. N. Lebedev Physical Institute of Russian Academy of Sciences in order to  support the data processing of Radioastron mission. This software has an implementation of the new delay calculation model for space-ground interferometry. The new model take into account the space-ground interferometer peculiarities such as relativistic and random drift of the onboard H-maser, absence of onboard clock. During the five years of Radioastron mission ASC Correlator proved to be a high efficient and reliable VLBI data processing instrument.

\section*{Acknowledgments}

Radioastron project is led by Astro Space Center of Lebedev Physical Institute of Russian Academy of Sciences and Lavochkin Scientific and Production Association under a contract with Russian Federal Space Agency in collaboration with partner organizations in Russia and other countries. 

Partly based on observations with the 100-m telescope of the MPIfR (Max-Planck-Institute for Radio Astronomy) at Effelsberg. The Arecibo Observatory is operated by SRI International under a cooperative agreement with the National Science Foundation (AST-1100968), and in alliance with Ana G. Mendez-Universidad Metropolitana, and the Universities Space Research Association. The National Radio Astronomy Observatory is a facility of the National Science Foundation operated under cooperative agreement by Associated Universities, Inc. Partly based on observations performed with radio telescopes of IAA RAS (Federal State Budget Scientific Organization Institue of Applied Astronomy of Russian Academy of Sciences).

We thank RadioAstron science teams for providing data being used in extensive RadioAstron ASC correlator testing and analysis presented in this paper.

We are grateful to Prof. M.~V.~Sazhin, K.~V.~Kuimov and Victor Zuga for useful discussions. We thank J. M. Anderson and P. A. Voitsik for productive discussions on the comparative analysis of ASC and DiFX correlators and anonymous referee for valuable comments and suggestions.

\end{document}